%% file: main.tex
\documentclass[aps,prl,reprint,superscriptaddress,longbibliography,floatfix]{revtex4-2}
\usepackage[T1]{fontenc}
\usepackage{amsmath,amssymb,bm,graphicx}
\usepackage{microtype}
\usepackage{xcolor}
\usepackage{balance}
\usepackage[colorlinks=true,linkcolor=blue!50!black,citecolor=blue!50!black,urlcolor=blue!50!black]{hyperref}
\hypersetup{pdftitle={Covariance Links Black Holes, Gravitational Collapse, and Cosmology},pdfsubject={Physical Review Letters manuscript}}
\newcommand{\dd}{\mathrm{d}}
\newcommand{\Q}{\mathcal Q}

\newcommand{\GZU}{School of Physics, Guizhou University, Guiyang 550025, China}
\newcommand{\BNU}{School of Physics and Astronomy, \mbox{Key Laboratory of Multiscale Spin Physics (Ministry of Education)}, Beijing Normal University, Beijing 100875, China}
\begin{document}
\title{Static Vacuum Geometry Encodes Gravitational Collapse and Cosmology}
\author{Yi Ling}
\email{lingy@ihep.ac.cn}
\affiliation{Institute of High Energy Physics,
Chinese Academy of Sciences, Beijing 100049, China}
\affiliation{School of Physics, University of Chinese Academy of Sciences, Beijing 100049, China}

\author{Jinsong Yang}
\email{jsyang@gzu.edu.cn}
\affiliation{\GZU}

\author{Zhangping Yu}
\email{yuzhangping@ucas.ac.cn}
\affiliation{International Centre for Theoretical Physics Asia-Pacific (ICTP-AP),
University of Chinese Academy of Sciences, Beijing 100190, China}

\author{Cong Zhang}
\email{cong.zhang@bnu.edu.cn}
\affiliation{\BNU}

\begin{abstract}
We show that, within covariant Hamiltonian  gravity in spherical symmetry, a prescribed one-parameter family of static vacuum geometries uniquely determines the local dynamics of gravitational collapse coupled to irrotational dust, despite the nonuniqueness of the underlying gravitational Hamiltonian. We further derive necessary and sufficient conditions under which the resulting dynamics admits a homogeneous and isotropic cosmological sector. These conditions establish a two-way correspondence between static vacuum geometries and modified Friedmann dynamics and impose a universal restriction on the spatial-curvature dependence of the latter. Our results provide a criterion for determining whether independently proposed black-hole and cosmological models can arise from a common covariant gravitational theory.
\end{abstract}
\maketitle

\emph{Introduction.---}
Understanding the quantum nature of gravity is 
one of the central challenges in fundamental physics. Since obtaining a complete description of quantum gravity remains difficult, considerable progress has been made through symmetry-reduced and effective models, which allow possible quantum gravity effects to be investigated in simplified settings. Quantum cosmology \cite{Bojowald:2001xe,Ashtekar:2006rx,Ashtekar:2006wn,Ashtekar:2011ni} and quantum-corrected black holes \cite{Dymnikova:1992ux,Ayon-Beato:1998hmi,Bronnikov:2000vy,Ayon-Beato:2000mjt,Modesto:2004xx,Hayward:2005gi,Gambini:2013ooa,Ashtekar:2018lag,Bodendorfer:2019cyv,Kelly:2020uwj} are two prominent examples of this approach. However, these two directions have largely developed independently, with quantum gravity modifications often introduced separately in each symmetry sector. Consequently, models that are individually consistent within their respective sectors need not be mutually consistent: there may be no common gravitational dynamics capable of reproducing both. Establishing consistency relations between these sectors is therefore important both for assessing their theoretical compatibility and for enabling observations in different gravitational settings to provide complementary constraints on quantum gravity.

Birkhoff’s theorem plays a crucial role in relating these two sectors. In general relativity, it guarantees that the spherically symmetric vacuum geometry is uniquely determined by the Schwarzschild mass. More generally, for a gravitational theory satisfying a Birkhoff-like theorem, its spherically symmetric vacuum solutions are characterized by a unique one-parameter family of static geometries \cite{Bueno:2025qjk,Cafaro:2024vrw}. On the other hand, in cosmology, homogeneity and isotropy reduce the spacetime geometry to a single scale factor, whose evolution is governed by a (modified) Friedmann equation.
Consequently, this theory determines both a unique family of static, spherically symmetric vacuum geometries and a modified Friedmann equation governing cosmological evolution. Although physically distinct, these two sectors cannot be specified independently if they are to arise from the same underlying gravitational dynamics. This raises a concrete question: Does the correspondence between the static vacuum geometry and the modified Friedmann equation depend on the details of the underlying gravitational theory, or can one be determined from the other independently of those details?

Significant progress has been made toward establishing the correspondence. In general relativity, Lema\^itre--Tolman--Bondi (LTB) dynamics describes spherical dust evolution \cite{Lemaitre:1933gd,Tolman:1934za,Bondi:1947fta}, reduces to Schwarzschild geometry in the vacuum sector, and contains Friedmann--Lema\^itre--Robertson--Walker (FLRW) cosmology as its homogeneous sector \cite{Friedman:1922kd,Robertson:1935zz,Walker:1937qxv}.
Beyond general relativity, the interplay between gravitational collapse, black-hole geometries, and cosmological dynamics has been extensively investigated~\cite{PhysRevD.109.064023,PhysRevD.110.104017,qrbb-mdvm,8f3j-zcxh,Lewandowski:2022zce,Giesel:2024mps,Ling:2025ncw}.
 In particular,  quantum-corrected black-hole geometries have been derived from modified cosmological dynamics by imposing additional assumptions on the junction conditions and the symmetries of the resulting black-hole geometries~\cite{Lewandowski:2022zce}. Conversely, modified FLRW dynamics has been derived from prescribed regular black-hole geometries within certain classes of polymerized and mimetic gravity models~\cite{Giesel:2024mps}. Similar black-hole--cosmology relations have also emerged in quasi-topological gravity~\cite{Ling:2025ncw,qrbb-mdvm}. Despite this progress, it remains unclear to what extent such connections follow from general principles rather than additional assumptions or specific choices of gravitational dynamics.

General covariance, a fundamental principle of modern physics, provides a natural starting point for addressing this issue. While general covariance can be made manifest in the Lagrangian formulation, its implementation in effective theories inspired by canonical quantum gravity is more subtle. These theories are naturally formulated within the Hamiltonian framework, where the foliation of spacetime into spatial hypersurfaces makes it nontrivial to preserve general covariance when introducing quantum corrections. Imposing this requirement therefore places strong restrictions on the allowed gravitational dynamics. In Refs.~\cite{Zhang:2024khj,Zhang:2024ney,Zhang:2025ccx}, a framework of covariant Hamiltonian gravity in spherical symmetry was developed, in which imposing general covariance, together with the requirement that a prescribed one-parameter family of static, spherically symmetric geometries be vacuum solutions, allows one to reconstruct gravitational dynamics without specifying a particular form of quantum gravity corrections. Remarkably, the prescribed family turns out to be the unique spherically symmetric vacuum solution family of the reconstructed dynamics, so that Birkhoff’s theorem is automatically satisfied. Nevertheless, the reconstruction itself is not unique: different generally covariant gravitational dynamics can admit exactly the same family of vacuum geometries. This ambiguity persists after coupling to dust, leading to distinct gravity-dust theories sharing the same vacuum geometries. Consequently, it remains unclear whether this ambiguity affects the correspondence between static vacuum geometries and cosmological dynamics.

In this Letter, we show that the ambiguity does not
prevent a definite correspondence between static vacuum
geometries and cosmological dynamics.
First, we demonstrate that, when coupled to irrotational
dust, all gravitational reconstructions of the same
prescribed static vacuum family yield identical physical LTB dynamics.
Thus, the static vacuum geometry uniquely determines the corresponding dust evolution.
Second, we derive necessary and sufficient conditions
for the resulting LTB dynamics to admit an exact FLRW
sector. For geometries satisfying these conditions, the static
metric functions and the modified Friedmann equation
are determined by the same function, establishing a
two-way correspondence.
Furthermore, this function must be independent of
the spatial curvature $k$, implying that the modified
Friedmann equations for $k=0,\pm1$ can differ only
through the curvature term $k/a^2$.
Our results thus provide consistency criteria for
relating black-hole geometries and cosmological models with different spatial curvatures within a generally covariant gravitational framework.

\emph{Dust dynamics from the static metric.---}Let us first briefly recall how gravitational dynamics is encoded in the Hamiltonian formulation. In general relativity, the $3+1$ decomposition splits the Einstein equations into projections normal and tangential to spatial hypersurfaces. The normal-normal and normal-tangential projections yield the Hamiltonian constraint $H$ and the diffeomorphism constraint $H_x$, respectively, while the remaining spatial projections correspond to Hamilton's equations governing the evolution of the gravitational variables. The evolution is generated by the Hamiltonian density $NH+N^xH_x$, where $N$ and $N^x$ denote the lapse function and shift vector, respectively, and act as Lagrange multipliers. Thus, the gravitational constraints, together with the canonical Poisson structure, completely encode the gravitational dynamics. 

In the reconstructed gravitational theories of Ref.~\cite{Zhang:2025ccx}, the phase-space structure, including the canonical variables and their Poisson brackets, as well as the diffeomorphism constraint $H_x$, is kept unchanged from classical general relativity. The modification induced by quantum gravity effects is entirely encoded in the Hamiltonian constraint, with $H$ replaced by an effective constraint $H_{\rm eff}$. Let us start from a prescribed one-parameter family of static, spherically symmetric metrics, which in the Schwarzschild-like coordinates $(t_s,r,\theta,\phi)$ reads
\begin{equation}
\dd s^2=-F(r;m)\dd t_s^2+G(r;m)^{-1}\dd r^2+r^2\dd\Omega^2.
\label{eq:static-family}
\end{equation}
Here $m$ is the mass parameter labeling the family of metrics, $\dd\Omega^2=\dd\theta^2+\sin^2(\theta)\dd\phi^2$ is the standard metric on 2-sphere. The reconstruction algorithm determines the allowed form of $H_{\rm eff}$ such that the prescribed geometries \eqref{eq:static-family} are vacuum solutions of the reconstructed gravitational dynamics.

In this Letter, we minimally couple the reconstructed gravitational theories to irrotational dust, described by the canonical pair $(T,P_T)$. In the Hamiltonian formulation, this amounts to adding the dust contributions $H_{\rm eff}^{\rm d}$ and $H_x^{\rm d}$ to the gravitational Hamiltonian and diffeomorphism constraints $H_{\rm eff}$ and $H_x$, respectively. The dynamics of the coupled system is determined by the constraint equations $H_{\rm eff}+H_{\rm eff}^{\rm d}=0$ and $H_x+H_x^{\rm d}=0$, together with Hamilton's equations generated by the Hamiltonian density $N(H_{\rm eff}+H_{\rm eff}^{\rm d})+N^x(H_x+H_x^{\rm d})$. This is analogous to introducing matter into the vacuum Einstein equations, where $G_{\mu\nu}=0$ is replaced by $G_{\mu\nu}-8\pi T_{\mu\nu}=0$.

To solve the dynamics, we choose the dust-comoving gauge $T(t,x)=t$ and $N^x=0$. Preservation of the gauge condition $T=t$ further fixes the lapse function to $N=1$, while the dust equation of motion
implies $\partial_tP_T=0$. The dust four-velocity is then given by $u^\mu=-g^{\mu\nu}\partial_\nu T=\delta^\mu_0$, implying that the dust particles follow trajectories of constant radial coordinate $x$, which therefore labels the individual spherical shells of the dust ball. In this gauge, the dynamics of the dust ball is characterized by the areal radius $R(t,x)$ of each spherical shell and the matter density $\rho(t,x)$, which describe the evolution of the shell radii and the density distribution, respectively. The evolution of the gravitational geometry, on the other hand, is encoded in the spacetime metric.

Solving the equations of motion of the dust-coupled system in the dust-comoving gauge (see appendix \ref{app:Meff-dynamics} for details), we obtain the generalized LTB metric
\begin{equation}
\dd s^2=-\dd t^2+
\frac{[\partial_xR(t,x)]^2}{e(x)^2\Q(R(t,x);\mathfrak m(x))}\dd x^2
+R(t,x)^2\dd\Omega^2,
\label{eq:ltb-metric}
\end{equation}
where the function $\Q$ is determined by the prescribed static metric~\eqref{eq:static-family}. More precisely, $\Q(x;m)=G(x;m)/F(x;m)$. The areal radius $R(t,x)$ of each spherical shell satisfies
\begin{equation}
(\partial_tR)^2
=e(x)^2\Q(R;\mathfrak m(x))
-G(R;\mathfrak m(x)).
\label{eq:shell}
\end{equation}
Here, $\mathfrak m(x)$ is a conserved mass function, while $e(x)$ is a conserved shell-energy profile related to $\mathfrak m(x)$ and the dust momentum $P_T(x)$ by
\begin{equation}\label{eq:einPT}
e(x)=\frac{\partial_x\mathfrak m(x)}{4\pi P_T(x)}.
\end{equation}
The matter density follows from the Hamiltonian constraint:
\begin{equation}
\rho(t,x)
=
\frac{\mathfrak m'(x)\sqrt{\Q(R(t,x);\mathfrak m(x))}}
{4\pi R(t,x)^2\partial_xR(t,x)}.
\label{eq:density}
\end{equation}
When $\mathfrak m(x)$ is spatially constant, the matter density vanishes, recovering the vacuum case.

Equations~\eqref{eq:ltb-metric}--\eqref{eq:density} give the general local solution to the equations of motion of the dust-coupled system wherever the comoving gauge is valid.
These equations show that, given a prescribed family of
static metrics~\eqref{eq:static-family}, characterized by
the functions $F$ and $G$, the corresponding LTB metric,
the evolution of the areal radius $R(t,x)$, and the dust
density $\rho(t,x)$ are completely determined by the
dust data $P_T(x)$ and $\mathfrak m(x)$, once the initial
radial coordinate and evolution branch are fixed.
Remarkably, these results are independent of the choice
of gravitational reconstruction.

The independence of the LTB dynamics from the gravitational reconstruction can be understood as follows (see Appendix~\ref{app:Meff-dynamics} for details). We first express the prescribed vacuum metric~\eqref{eq:static-family} in coordinates $(t,x)$ with $N=1$ and $N^x=0$, and then promote the constant mass parameter $m$ to a spacetime-dependent function $\mathfrak m(t,x)$. The generality of this construction is established in
Appendix~\ref{app:Meff-dynamics}.
Substituting this ansatz into the coupled constraint and evolution equations, we find that each equation takes the schematic form $
a_0+a_x\partial_x\mathfrak m+a_t\partial_t\mathfrak m=s,$
where $s$ denotes the dust source term. Crucially, $a_0$ vanishes identically for the vacuum-inspired ansatz. To see this, we note that $a_0$ contains neither derivatives of $\mathfrak m$ nor dust contributions. Thus, when $\mathfrak m$ is constant and the dust source vanishes, the equations reduce to $a_0=0$. Since the prescribed metric is a vacuum solution for any constant mass parameter $m$, we have $a_0=0$ for arbitrary values of $\mathfrak m$. As $a_0$ contains no derivatives of $\mathfrak m$, it remains identically zero even when $\mathfrak m$ depends on spacetime coordinates. The coupled equations therefore reduce to $a_x\partial_x\mathfrak m+a_t\partial_t\mathfrak m=s$. As demonstrated in Appendix~\ref{app:Meff-dynamics}, all dependence on the reconstruction freedom cancels from these reduced equations, yielding the LTB dynamics given by Eqs.~\eqref{eq:ltb-metric}--\eqref{eq:density} for all admissible reconstructions. In particular, the reduced equations imply $\partial_t\mathfrak m=0=\partial_t e$, showing that both the mass function and the shell-energy profile are conserved along each comoving dust shell.

\emph{Cosmological selection.---}According to the results of the previous section, the dynamics is determined once the dust configuration, characterized by two functions $e(x)$ and $\mathfrak m(x)$, is specified. We now ask whether such a configuration can be chosen so that the resulting dynamics admits an FLRW sector. More precisely, we ask whether there exist functions $e(x)$, or equivalently $P_T(x)$, and $\mathfrak m(x)$ for which the metric~\eqref{eq:ltb-metric} reduces to an FLRW metric while the dust density~\eqref{eq:density} remains homogeneous throughout the evolution, i.e., depends only on $t$.

Requiring the metric in Eq.~\eqref{eq:ltb-metric} to reduce to an FLRW metric, one gets $R(t,x)=a(t)x$ and $e(x)^2\Q(a(t)x;\mathfrak m(x))=1-kx^2$, where $a(t)$ is the cosmological scale factor and $k=0,\pm1$ denotes the spatial curvature. Since the right-hand side of the latter relation is independent of $t$, $\Q$ must take the form $\Q(R,m)=q(m)$. Requiring the density~\eqref{eq:density} to be homogeneous then gives $\partial_x\mathfrak m(x)\sqrt{q(\mathfrak m(x))}=\alpha x^2$, with $\alpha$ a constant. This relation, together with $e(x)^2q(\mathfrak m(x))=1-kx^2$, will be referred to as the \emph{FLRW configuration conditions}. They determine the dust configuration $\mathfrak m(x)$ and $e(x)$ once $q(m)$ is specified.

These conditions ensure that the metric and density can take the FLRW form. It remains to require that the ansatz $R(t,x)=a(t)x$ be preserved by the dynamics. We therefore impose the shell equation~\eqref{eq:shell}. Using the results derived above, it becomes 
\begin{equation}\label{eq:friedmann0}
H^2+\frac{k}{a^2}=\frac{1-G\bigl(a x,\mathfrak m(x)\bigr)}{a^2x^2},
\end{equation}
 where $H=\dot a/a$ is the Hubble parameter. For the ansatz $R(t,x)=a(t)x$ to be preserved under evolution, the right-hand side must therefore be independent of $x$ for $\mathfrak m(x)$ fixed by the FLRW configuration conditions. This requirement restricts $G$ to the form (see Appendix \ref{app:selection})
\begin{equation}
G(R,m)=
1-R^2 g\left(\frac{\mathcal M(m)}{R^3}\right),
\label{eq:general-selection}
\end{equation}
for some function $g$, where $\mathcal M(m)$ is determined by $\partial_m\mathcal M(m)=\sqrt{q(m)}$ and $\mathcal M(\mathfrak m(0))=0$. Then, Eq.~\eqref{eq:density} together with $R(t,x)=a(t)x$ gives $\partial_x\mathcal M(\mathfrak m(x))=4\pi\rho a^3x^2$. Since the FLRW configuration conditions imply that $\rho$ depends only on $t$, integrating this relation yields $\mathcal M(\mathfrak m(x))=4\pi\rho a^3x^3/3$. The evolution of the corresponding FLRW scale factor is therefore governed by
\begin{equation}
H^2+\frac{k}{a^2}
=
g\left(\frac{4\pi\rho}{3}\right),
\quad
\dot\rho+3H\rho=0,
\label{eq:friedmann}
\end{equation}
where the first equation, obtained from Eq.~\eqref{eq:friedmann0}, is the modified Friedmann equation, while the second follows from $\partial_t\mathcal M(\mathfrak m(x))=0$ and expresses energy conservation.

We now summarize the restriction on the static vacuum
geometry imposed by the existence of an FLRW sector.
Returning to Schwarzschild-like coordinates,
we find that the one-parameter family of static,
spherically symmetric vacuum metrics~\eqref{eq:static-family}
admits an FLRW sector if and only if its metric functions
can be written as
\begin{equation}\label{eq:vacuumFLRW}
F(r;m)=G(r;m)
=1-r^2g\left(\frac{\mathcal M(m)}{r^3}\right).
\end{equation}
Here, we note that $F=G$ is achieved by rescaling the static time coordinate by a constant factor $1/\partial_m\mathcal M(m)$ for each fixed $m$, leaving the spacetime geometry unchanged. 

An important consequence is that the same function $g$ must govern the modified Friedmann equations for
different spatial curvatures $k$.
Thus, within the present framework, the equations for
$k=0,\pm1$ can differ only through the term $k/a^2$.
In particular, specifying the modified Friedmann equation for $k=0$ uniquely determines those for $k=\pm1$. This provides a nontrivial consistency condition for
quantum cosmological models constructed independently
for different spatial curvatures.

\emph{Examples.---}For fixed $\ell>0$, the Bardeen family has $F_{\rm B}=G_{\rm B}=1-2mr^2/(r^2+\ell^2)^{3/2}$ \cite{Ayon-Beato:2000mjt}. Since $[1-G_{\rm B}]/r^2$ is not a function of $m/r^3$, the Bardeen family does not admit an exact FLRW dust sector within the present framework.

The Hayward family, $F_{\rm H}=G_{\rm H}=1-2mr^2/(r^3+2m\ell^2)$ \cite{Hayward:2005gi}, satisfies Eq.~\eqref{eq:vacuumFLRW} with $g(z)=2z/(1+2\ell^2z)$, yielding
\begin{equation}
H^2+\frac{k}{a^2}
=\frac{8\pi\rho/3}{1+8\pi\ell^2\rho/3}.
\label{eq:hayward-friedmann}
\end{equation}
For $k=0$, the high-density limit $H^2\to\ell^{-2}$ shows that the regular black-hole core and the limiting cosmological expansion rate are controlled by the same scale.

Our consistency condition also places restrictions on effective loop quantum cosmology (LQC). Several existing effective models with $k=\pm1$ contain curvature-dependent quantum corrections beyond the term $k/a^2$~\cite{PhysRevD.75.024035,PhysRevD.75.023523,Yang:2022aec}, which cannot be accommodated by the same function $g$ governing the standard flat LQC model~\cite{Ashtekar:2006rx}. This suggests that these effective LQC models with different spatial curvatures cannot all arise as FLRW sectors of a single gravitational theory within the present covariant Hamiltonian gravity.

\emph{Conclusion.---} Previous studies have established related connections between black-hole geometries and cosmological dynamics through junction conditions~\cite{Lewandowski:2022zce}, within specific classes of polymerized and mimetic gravity models~\cite{Giesel:2024mps}, and in higher-dimensional quasi-topological gravity~\cite{Ling:2025ncw}. These studies provide important realizations of such connections through different approaches. Our work goes further in two respects.

First, we demonstrate that, within covariant Hamiltonian gravity in spherical symmetry, all admissible gravitational reconstructions of the same prescribed family of static vacuum geometries yield identical LTB dynamics when minimally coupled to irrotational dust. Thus, the dynamics of gravitational collapse is uniquely determined by the static vacuum geometry, independently of the gravitational reconstruction ambiguity.

Second, we derive necessary and sufficient conditions for the resulting dust dynamics to admit an exact FLRW sector. These conditions restrict the admissible static vacuum geometries and establish a correspondence between their metric functions and the modified Friedmann equation. In particular, the modified Friedmann equations for different spatial curvatures must be governed by the same function, differing only by the classical curvature term. This provides a consistency criterion for assessing whether independently constructed quantum-corrected black-hole and cosmological models can arise from a common covariant gravitational theory.

 Although our analysis is carried out within covariant Hamiltonian gravity in spherical symmetry, the universality found here may reflect more general principles. In vacuum, Birkhoff's theorem ensures that the entire spherically symmetric solution space is characterized by a one-parameter family labeled by $m$. As shown in Ref.~\cite{Zhang:2025ccx}, this structure is encoded in the Hamiltonian constraint, with $m$ arising as an integration constant. Upon coupling to dust, the gravitational part of the Hamiltonian constraint remains unchanged, while the dust contribution enters as a source term. Solving the equations of motion then promotes the integration constant in vacuum to a function of the radial coordinate $x$, $m\to m(x)$, allowing the dynamics of individual dust shells to be determined directly from the vacuum solution. General covariance, on the other hand, restricts the admissible gravitational dynamics and enables their reconstruction from the prescribed static vacuum geometry. These observations suggest that the universality found here may have a more general origin in the interplay between Birkhoff's theorem and general covariance, rather than in the specific form of the underlying gravitational dynamics.

\begin{acknowledgments}
This work was supported by the National Natural Science
Foundation of China (NSFC) under Grant Nos. 12165005,
12275275, 12505055, and 12675082, and by the
Fundamental Research Funds for the Central Universities.
\end{acknowledgments}

\bibliography{references}

\onecolumngrid
\vspace{5mm}
\begin{center}
{\large\bfseries Appendices}
\end{center}
\vspace{5mm}
\twocolumngrid
\appendix

\setcounter{secnumdepth}{1}
\input{appendix}

\end{document}

%% file: appendix.tex
\section{Covariant Hamiltonian and reconstruction}
\label{app:reconstruction}
As discussed in Refs.~\cite{Zhang:2024khj,Zhang:2024ney}
and references therein, the gravitational phase space in
spherical symmetry is described by two canonical pairs
$(K_I,E^I)$, with the nonvanishing Poisson brackets
$\{K_1(x),E^1(y)\}=2\delta(x-y)$ and
$\{K_2(x),E^2(y)\}=\delta(x-y)$.
To express the effective Hamiltonian constraint $H_{\rm eff}$,
it is convenient to introduce five basic scalars:
\begin{equation*}
s_1=E^1,\
s_2=K_2,\
s_3=\frac{K_1}{E^2},\
s_4=\frac{\partial_x s_1}{E^2},\
s_5=\frac{\partial_x s_4}{E^2}.
\end{equation*}
The classical Hamiltonian constraint of general relativity can be expressed in terms of these scalars, up to an overall density factor. In Refs.~\cite{Zhang:2024khj,Zhang:2024ney,Zhang:2025ccx}, the effective Hamiltonian constraint is constructed from the same scalars, but with a more general functional dependence. 

A necessary condition for covariance is the closure of the constraint algebra. In Refs.~\cite{Zhang:2024khj,Zhang:2025ccx}, one assumed
\begin{equation}\label{eq:poisson}
\{H_{\rm eff}[N],H_{\rm eff}[M]\}
=
H_x\left[S\left(N\partial_xM-M\partial_xN\right)\right],
\end{equation}
where $S=\mu E^1/(E^2)^2$ is the structure function and $H[f]\equiv\int Hf\,\dd x$. The deformation factor $\mu$ reduces to unity in classical general relativity but may depend on the phase-space variables in the effective theory. Closure of the constraint algebra alone is not sufficient for spacetime covariance. A covariant spacetime interpretation further requires the same deformation factor $\mu$ to enter the spacetime metric,
\begin{equation}\label{eq:effmetric}
\dd s^2=-N^2\dd t^2+\frac{(E^2)^2}{\mu E^1}(\dd x+N^x\dd t)^2+E^1\dd\Omega^2,
\end{equation}
where $N$ and $N^x$ are the lapse function and shift vector, respectively. General covariance requires that the metrics in Eq.\eqref{eq:effmetric} corresponding to different choices of $N$ and $N^x$ be related by spacetime diffeomorphisms. Together with the closure condition \eqref{eq:poisson}, this requirement fixes the effective Hamiltonian constraint to take the form 
\begin{equation*}
H_{\rm eff}=-2E^2\Big[\partial_{s_1}M_{\rm eff}
+\frac{\partial_{s_2}M_{\rm eff}}{2}s_3
+\frac{\partial_{s_4}M_{\rm eff}}{s_4}s_5
+\mathcal R\Big],
\end{equation*}
where $\mathcal R$ is a function of $s_1$ and the mass function $M_{\rm eff}$, while $M_{\rm eff}$ depends on $s_1$, $s_2$, and $s_4$ and is constrained by
\begin{equation}\label{eq:appendix-mass-covariance}
\big(\partial_{s_2}M_{\rm eff}\big)^2
=
\frac{\mu s_1}{4}\big((s_4)^2+\mathcal Z\big).
\end{equation}
Here $\mathcal Z=\mathcal Z(s_1,M_{\rm eff})$ is arbitrary. Moreover, general covariance further requires that $\mu$ can only depend on $s_1$ and $M_{\rm eff}$. A covariant theory within this framework is therefore characterized by three functions $\mathcal R$, $\mu$, and $\mathcal Z$. 

To reconstruct the dynamics from a prescribed family of static, spherically symmetric vacuum geometries, one needs to first specify the on-shell relation $$M_{\rm eff}=f(E^1,m),$$
where $\partial_mf\neq 0$ is required. 
Once the function $f$ is chosen, the reconstruction algorithm of Ref.~\cite{Zhang:2025ccx} gives
\begin{equation}\label{eq:appendix-mu}
\mathcal R=-\partial_{E^1}f,\qquad
\mu=\frac{\Q}{(\partial_mf)^{2}},\qquad
\mathcal Z=-\frac{4G}{\mu}.
\end{equation}
Here $G(\sqrt{E^1},m)$ is the function that appears in the prescribed metric \eqref{eq:static-family}, and $\Q(\sqrt{E^1},m)=G(\sqrt{E^1},m)/F(\sqrt{E^1},m)$. These expressions are first obtained in terms of $E^1$ and $m$, and are then rewritten as functions of $(s_1,M_{\rm eff})$ by expressing $E^1$ and $m$ in terms of $s_1$ and $M_{\rm eff}$ using $s_1=E^1$ and $M_{\rm eff}=f(E^1,m)$.

The phase space of irrotational dust is described by the canonical pair $(T,P_T)$ with the Poisson bracket $\{T(x),P_T(y)\}=\delta(x-y)/(4\pi)$. Its contributions to the Hamiltonian and diffeomorphism constraints are
\begin{equation*}
H^{\rm d}_{\rm eff}
=
4\pi P_T\sqrt{1+S(\partial_xT)^2},
\quad
H^{\rm d}_x
=
4\pi P_T\partial_xT,
\end{equation*}
where $S=\mu E^1/(E^2)^2$, and we have restricted to the positive-energy dust sector with $P_T>0$. 
They retain their classical form, with the classical spatial metric replaced by that associated with the effective metric in Eq.~\eqref{eq:effmetric}. It turns out in \cite{Yang:2025ufs} that this form of $H_{\rm eff}^{\rm d}$ and $H_x^{\rm d}$ ensures general covariance for the dust coupled model. 

The total constraints are therefore $H^{\rm tot}=H_{\rm eff}+H^{\rm d}_{\rm eff}$ and $H_x^{\rm tot}=H_x+H_x^{\rm d}$.

\section{Solving the dynamics}\label{app:Meff-dynamics}

In vacuum, imposing $N=1$ and $N^x=0$, the constraint
and evolution equations can be written as
\begin{equation}\label{eq:constraintM}
\begin{aligned}
\partial_xM_{\rm eff}
=&-\mathcal R\partial_xE^1,\\
\partial_tM_{\rm eff}
=&-\mathcal R\partial_tE^1,\\
(\partial_tE^1)^2
=&\mu E^1
\left[
\left(\frac{\partial_xE^1}{E^2}\right)^2
+\mathcal Z
\right],\\
\partial_t\left(\frac{\partial_xE^1}{E^2}\right)
=&-\frac{\partial_xE^1}{E^2}
\partial_tE^1\,
\partial_{M_{\rm eff}}\mathcal R.
\end{aligned}
\end{equation}
By virtue of the reconstruction relations~\eqref{eq:appendix-mu}, which relate $\mathcal R$, $\mu$, and $\mathcal Z$ to the prescribed static metric, the general solution to Eq.~\eqref{eq:constraintM} is given by (see~\cite{Zhang:2025ccx} for a proof of its generality)
\begin{equation}\label{eq:solvvacuum}
M_{\rm eff}=f(E^1,m),\
\frac{\partial_xE^1}{E^2}=2e\partial_mf,\
E^1=R^2,
\end{equation}
where $m$ is a constant, $e$ is an arbitrary function that depends only on $x$, and $R=R(t,x)$ satisfies
\begin{equation}
(\partial_tR)^2=e(x)^2\mathcal Q(R;m)-G(R;m).
\end{equation}
Here, we need to use the following useful relation derived from
Eq.~\eqref{eq:appendix-mu}:
\begin{equation}\label{eq:keyrelation}
\partial_{M_{\rm eff}}\mathcal R\partial_mf
=-\partial_m\partial_{E^1}f,
\end{equation}
where both sides are evaluated at $E^1=R^2$ and
$M_{\rm eff}=f(R^2,m)$.

When dust is coupled, we adopt the comoving gauge
$T=t$ and $N^x=0$. Preservation of the condition
$T=t$ fixes $N=1$, while the dust equation of motion
implies $\partial_tP_T=0$.
In this gauge, the complete set of equations of motion reads as
\begin{equation}\label{eq:constraintMdust}
\begin{aligned}
\partial_xM_{\rm eff}=&-
\partial_xE^1\,\mathcal R+\frac{2\pi P_T}{E^2}
\partial_xE^1,\\
\partial_tM_{\rm eff}=&-\partial_tE^1\,\mathcal R,\\
(\partial_tE^1)^2
=&
\mu E^1
\left[
\left(\frac{\partial_xE^1}{E^2}
\right)^2
+\mathcal Z
\right],\\
\partial_t\left(\frac{\partial_xE^1}{E^2}\right)=&-\frac{\partial_xE^1}{E^2}\partial_tE^1\partial_{M_{\rm eff}}\mathcal{R}.
\end{aligned}
\end{equation}
Compared with the vacuum equations in ~\eqref{eq:constraintM}, the only difference is the additional dust contribution involving $P_T(x)$ in the first equation. 

To solve Eq.~\eqref{eq:constraintMdust}, let us parameterize
$M_{\rm eff}$ as
\begin{equation}
M_{\rm eff}=f(E^1,\mathfrak m(t,x)),
\end{equation}
where the constant mass parameter $m$ in the vacuum
solution is promoted to a spacetime-dependent function
$\mathfrak m(t,x)$.
Then, the first two equations in Eq.~\eqref{eq:constraintMdust}
give
\begin{equation}
\begin{aligned}
\partial_xE^1(\partial_{E^1}f+\mathcal R)
+\partial_mf\,\partial_x\mathfrak m
&=2\pi P_T\frac{\partial_xE^1}{E^2},\\
\partial_tE^1(\partial_{E^1}f+\mathcal R)
+\partial_mf\,\partial_t\mathfrak m&=0.
\end{aligned}
\end{equation}
As discussed in the main text, these equations take the
schematic form
$a_0+a_x\partial_x\mathfrak m
+a_t\partial_t\mathfrak m=s$.
In particular, the terms independent of the derivatives
of $\mathfrak m$ vanish identically due to
$\partial_{E^1}f+\mathcal R=0$ from
Eq.~\eqref{eq:appendix-mu}, 
reflecting the fact that
the prescribed metric satisfies the vacuum equations
for arbitrary constant $m$, as discussed in the main text.
Consequently, the equations reduce to
\begin{equation}\label{eq:solvMeffdust}
\begin{aligned}
\partial_mf\partial_x\mathfrak m
&=2\pi P_T\frac{\partial_xE^1}{E^2},\\
\partial_mf\,\partial_t\mathfrak m&=0.
\end{aligned}
\end{equation}
Since $\partial_mf\neq0$, the second equation immediately
implies
\begin{equation}\label{eq:mindependt}
\partial_t\mathfrak m=0,
\end{equation}
where the reconstruction-dependent factor $\partial_m f$ has canceled, as discussed in the main text.

Using Eq.~\eqref{eq:mindependt} and the relation
\eqref{eq:keyrelation}, the last equation in
Eq.~\eqref{eq:constraintMdust} simplifies to
\begin{equation}
\partial_t\left(\frac{\partial_xE^1}{E^2}\right)
=
\frac{\partial_xE^1}{E^2}
\frac{\partial_t(\partial_m f)}{\partial_m f}.
\end{equation}
Here, $\partial_m f$ is a function of $E^1$ and
$\mathfrak m$, and we have used
$\partial_t\mathfrak m=0$ to obtain
$\partial_t(\partial_m f)
=(\partial_{E^1}\partial_m f)\partial_tE^1$. 
Integrating this equation with respect to time, we obtain
\begin{equation}
\frac{\partial_xE^1}{E^2}=2e\partial_m f,
\end{equation}
where $e$ is an arbitrary function depending only on $x$.
This recovers the second relation in Eq.~\eqref{eq:solvvacuum}, with the constant mass parameter $m$ replaced by the function $\mathfrak m(x)$. Substituting this result into Eq.~\eqref{eq:solvMeffdust},
we obtain
\begin{equation}
\partial_x\mathfrak m=4\pi e P_T,
\end{equation}
where the reconstruction-dependent factor $\partial_m f$
cancels out once again. Finally, setting $E^1=R^2$ and using the reconstruction relations in Eq.~\eqref{eq:appendix-mu}, which hold pointwise for arbitrary values of the mass parameter, the third equation in Eq.~\eqref{eq:constraintMdust} yields Eq.~\eqref{eq:shell}.

Since the parameterization $M_{\rm eff}=f(E^1,\mathfrak m)$ is locally invertible for $\partial_mf\neq0$, and no further restrictions have been imposed in solving the complete set of equations, the resulting LTB dynamics represents the general local solution in the comoving gauge.

\section{Derivation of the cosmological selection rule}
\label{app:selection}

From the results obtained in the main text, the existence of an FLRW sector requires $\Q(R,m)=q(m)$ and
\begin{equation}\label{eq:determinme}
\begin{aligned}
\partial_x\mathfrak m(x)\sqrt{q(\mathfrak m(x))}
=&
\alpha x^2,\\
e(x)^2q(\mathfrak m(x))=&1-kx^2.
\end{aligned}
\end{equation}
These equations, referred to as the FLRW configuration conditions, determine the dust configuration $\mathfrak m(x)$ and $e(x)$ once $q(m)$ is specified.

Conversely, suppose that $\Q(R,m)=q(m)$. Choosing $\mathfrak m(x)$ and $e(x)$ according to Eq.~\eqref{eq:determinme}, the ansatz $R(t,x)=a(t)x$ makes the metric~\eqref{eq:ltb-metric} FLRW and the density~\eqref{eq:density} homogeneous.

The remaining question is whether the form $R(t,x)=a(t)x$ is preserved by the dynamics. We therefore impose the shell equation~\eqref{eq:shell}. With $R(t,x)=a(t)x$, it becomes
\begin{equation}
H^2+\frac{k}{a^2}
=
\frac{1-G(a x,\mathfrak m(x))}{a^2x^2}
:=
Y(a x,\mathfrak m(x)),
\end{equation}
where $H=\dot a/a$. For the ansatz $R(t,x)=a(t)x$ to be preserved under evolution, the right-hand side must be independent of $x$, and hence
\begin{equation}\label{eq:conditionY}
\partial_xY(a x,\mathfrak m(x))=0.
\end{equation}

Now let us introduce $\widetilde{\mathcal M}$ by $\partial_m\widetilde{\mathcal M}=\sqrt{q(m)}$, and define $\widetilde{\mathcal M}(x)=\widetilde{\mathcal M}(\mathfrak m(x))$. Locally, the definition of $\widetilde{\mathcal M}(x)$ allows us to express $\mathfrak m(x)$ in terms of $\widetilde{\mathcal M}(x)$ and regard $Y$ as a function of $R=ax$ and $\widetilde{\mathcal M}$. Then, the condition~\eqref{eq:conditionY} becomes
\begin{equation}\label{eq:solveY}
R\partial_RY
+3(\widetilde{\mathcal M}-\widetilde{\mathcal M}_0)
\partial_{\widetilde{\mathcal M}}Y
=0,
\end{equation}
where we have used Eq.~\eqref{eq:determinme}, which gives
\begin{equation}
\widetilde{\mathcal M}(\mathfrak m(x))
=
\frac{\alpha}{3}x^3+\widetilde{\mathcal M}_0.
\end{equation}
The general solution to Eq.~\eqref{eq:solveY} is therefore
\begin{equation}
Y(R,m)
=
g\left(
\frac{\widetilde{\mathcal M}(m)-\widetilde{\mathcal M}_0}{R^3}
\right),
\end{equation}
for an arbitrary function $g$. We then define $\mathcal M$ by
\begin{equation}
\mathcal M(m)
=
\widetilde{\mathcal M}(m)-\widetilde{\mathcal M}_0,
\end{equation}
so that $\partial_m\mathcal M(m)=\sqrt{q(m)}$ and $\mathcal M(\mathfrak m(0))=0$, in agreement with the definition used in the main text. Hence,
\begin{equation}\label{eq:app-Gselection}
G(R,m)
=
1-R^2g\left(
\frac{\mathcal M(m)}{R^3}
\right).
\end{equation}
Together with $\Q(R,m)=q(m)$, Eq.~\eqref{eq:app-Gselection} is sufficient for the dust dynamics to admit an FLRW sector.

%% file: references.bib
@article{Yang:2022aec,
    author        = {Yang, Jinsong and Zhang, Cong and Zhang, Xiangdong},
    title         = {{Alternative k=-1 loop quantum cosmology}},
    eprint        = {2212.05748},
    archiveprefix = {arXiv},
    primaryclass  = {gr-qc},
    doi           = {10.1103/PhysRevD.107.046012},
    journal       = {Phys. Rev. D},
    volume        = {107},
    number        = {4},
    pages         = {046012},
    year          = {2023}
}

@article{8f3j-zcxh,
    title     = {Regular black hole formation in four-dimensional nonpolynomial gravities},
    author    = {Bueno, Pablo and Cano, Pablo A. and Hennigar, Robie A. and Murcia, {\'A}ngel J.},
    journal   = {Phys. Rev. D},
    volume    = {113},
    issue     = {2},
    pages     = {024019},
    numpages  = {16},
    year      = {2026},
    month     = {Jan},
    publisher = {American Physical Society},
    doi       = {10.1103/8f3j-zcxh},
    url       = {https://link.aps.org/doi/10.1103/8f3j-zcxh}
}

@article{qrbb-mdvm,
    title     = {Regular black holes from Oppenheimer-Snyder collapse},
    author    = {Bueno, Pablo and Cano, Pablo A. and Hennigar, Robie A. and Murcia, {\'A}ngel J. and Vicente-Cano, Aitor},
    journal   = {Phys. Rev. D},
    volume    = {112},
    issue     = {6},
    pages     = {064039},
    numpages  = {18},
    year      = {2025},
    month     = {Sep},
    publisher = {American Physical Society},
    doi       = {10.1103/qrbb-mdvm},
    url       = {https://link.aps.org/doi/10.1103/qrbb-mdvm}
}

@article{PhysRevD.109.064023,
    title     = {Nonsingular collapse of a spherical dust cloud},
    author    = {Alonso-Bardaji, Asier and Brizuela, David},
    journal   = {Phys. Rev. D},
    volume    = {109},
    issue     = {6},
    pages     = {064023},
    numpages  = {10},
    year      = {2024},
    month     = {Mar},
    publisher = {American Physical Society},
    doi       = {10.1103/PhysRevD.109.064023},
    url       = {https://link.aps.org/doi/10.1103/PhysRevD.109.064023}
}

@article{PhysRevD.110.104017,
    title     = {Embedding generalized Lema\^{\i}tre-Tolman-Bondi models in polymerized spherically symmetric spacetimes},
    author    = {Giesel, Kristina and Liu, Hongguang and Rullit, Eric and Singh, Parampreet and Weigl, Stefan Andreas},
    journal   = {Phys. Rev. D},
    volume    = {110},
    issue     = {10},
    pages     = {104017},
    numpages  = {29},
    year      = {2024},
    month     = {Nov},
    publisher = {American Physical Society},
    doi       = {10.1103/PhysRevD.110.104017},
    url       = {https://link.aps.org/doi/10.1103/PhysRevD.110.104017}
}

@article{Cafaro:2024vrw,
    author        = {Cafaro, Luca and Lewandowski, Jerzy},
    title         = {{Status of Birkhoff{\textquoteright}s theorem in the polymerized semiclassical regime of loop quantum gravity}},
    eprint        = {2403.01910},
    archiveprefix = {arXiv},
    primaryclass  = {gr-qc},
    doi           = {10.1103/PhysRevD.110.024072},
    journal       = {Phys. Rev. D},
    volume        = {110},
    number        = {2},
    pages         = {024072},
    year          = {2024}
}

@article{Bueno:2025qjk,
    author        = {Bueno, Pablo and Hennigar, Robie A. and Murcia, {\'A}ngel J.},
    title         = {{Birkhoff implies quasi-topological}},
    eprint        = {2510.25823},
    archiveprefix = {arXiv},
    primaryclass  = {gr-qc},
    doi           = {10.1088/1361-6382/ae6600},
    journal       = {Class. Quant. Grav.},
    volume        = {43},
    number        = {9},
    pages         = {095020},
    year          = {2026}
}

@article{PhysRevD.75.024035,
    title     = {Loop quantum cosmology of $k=1$ FRW models},
    author    = {Ashtekar, Abhay and Pawlowski, Tomasz and Singh, Parampreet and Vandersloot, Kevin},
    journal   = {Phys. Rev. D},
    volume    = {75},
    issue     = {2},
    pages     = {024035},
    numpages  = {26},
    year      = {2007},
    month     = {Jan},
    publisher = {American Physical Society},
    doi       = {10.1103/PhysRevD.75.024035},
    url       = {https://link.aps.org/doi/10.1103/PhysRevD.75.024035}
}

@article{PhysRevD.75.023523,
    title     = {Loop quantum cosmology and the $k=\ensuremath{-}1$ Robertson-Walker model},
    author    = {Vandersloot, Kevin},
    journal   = {Phys. Rev. D},
    volume    = {75},
    issue     = {2},
    pages     = {023523},
    numpages  = {13},
    year      = {2007},
    month     = {Jan},
    publisher = {American Physical Society},
    doi       = {10.1103/PhysRevD.75.023523},
    url       = {https://link.aps.org/doi/10.1103/PhysRevD.75.023523}
}

@article{Ling:2025ncw,
    author        = {Ling, Yi and Yu, Zhangping},
    title         = {{Big bounce and black bounce in quasi-topological gravity}},
    eprint        = {2509.00137},
    archiveprefix = {arXiv},
    primaryclass  = {gr-qc},
    doi           = {10.1088/1475-7516/2026/03/004},
    journal       = {JCAP},
    volume        = {03},
    pages         = {004},
    year          = {2026}
}

@article{Zhang:2024khj,
    author        = {Zhang, Cong and Lewandowski, Jerzy and Ma, Yongge and Yang, Jinsong},
    title         = {{Black holes and covariance in effective quantum gravity}},
    eprint        = {2407.10168},
    archiveprefix = {arXiv},
    primaryclass  = {gr-qc},
    doi           = {10.1103/PhysRevD.111.L081504},
    journal       = {Phys. Rev. D},
    volume        = {111},
    number        = {8},
    pages         = {L081504},
    year          = {2025}
}

@article{Zhang:2025ccx,
    author        = {Zhang, Cong and Cao, Zhoujian},
    title         = {{Covariant Dynamics from Static Spherically Symmetric Geometries}},
    eprint        = {2506.09540},
    archiveprefix = {arXiv},
    primaryclass  = {gr-qc},
    doi           = {10.1103/6sws-hfj7},
    journal       = {Phys. Rev. Lett.},
    volume        = {135},
    number        = {26},
    pages         = {261401},
    year          = {2025}
}

@article{Giesel:2024mps,
    author        = {Giesel, Kristina and Liu, Hongguang and Singh, Parampreet and Weigl, Stefan Andreas},
    title         = {{Regular black holes and their relationship to polymerized models and mimetic gravity}},
    eprint        = {2405.03554},
    archiveprefix = {arXiv},
    primaryclass  = {gr-qc},
    doi           = {10.1103/PhysRevD.111.064064},
    journal       = {Phys. Rev. D},
    volume        = {111},
    number        = {6},
    pages         = {064064},
    year          = {2025}
}

@article{Lewandowski:2022zce,
    author        = {Lewandowski, Jerzy and Ma, Yongge and Yang, Jinsong and Zhang, Cong},
    title         = {{Quantum Oppenheimer-Snyder and Swiss Cheese Models}},
    eprint        = {2210.02253},
    archiveprefix = {arXiv},
    primaryclass  = {gr-qc},
    doi           = {10.1103/PhysRevLett.130.101501},
    journal       = {Phys. Rev. Lett.},
    volume        = {130},
    number        = {10},
    pages         = {101501},
    year          = {2023}
}

@article{Hayward:2005gi,
    author        = {Hayward, Sean A.},
    title         = {{Formation and evaporation of regular black holes}},
    eprint        = {gr-qc/0506126},
    archiveprefix = {arXiv},
    doi           = {10.1103/PhysRevLett.96.031103},
    journal       = {Phys. Rev. Lett.},
    volume        = {96},
    pages         = {031103},
    year          = {2006}
}

@article{Ayon-Beato:2000mjt,
    author        = {Ayon-Beato, Eloy and Garcia, Alberto},
    title         = {{The Bardeen model as a nonlinear magnetic monopole}},
    eprint        = {gr-qc/0009077},
    archiveprefix = {arXiv},
    doi           = {10.1016/S0370-2693(00)01125-4},
    journal       = {Phys. Lett. B},
    volume        = {493},
    pages         = {149--152},
    year          = {2000}
}

@article{Zhang:2024ney,
    author        = {Zhang, Cong and Lewandowski, Jerzy and Ma, Yongge and Yang, Jinsong},
    title         = {{Black holes and covariance in effective quantum gravity: A solution without Cauchy horizons}},
    eprint        = {2412.02487},
    archiveprefix = {arXiv},
    primaryclass  = {gr-qc},
    doi           = {10.1103/d6ks-d576},
    journal       = {Phys. Rev. D},
    volume        = {112},
    number        = {4},
    pages         = {044054},
    year          = {2025}
}

@article{Yang:2025ufs,
    author        = {Yang, Jinsong and Zhang, Cong and Ma, Yongge},
    title         = {{Covariant effective spacetimes of spherically symmetric electrovacuum with a cosmological constant}},
    eprint        = {2503.15157},
    archiveprefix = {arXiv},
    primaryclass  = {gr-qc},
    doi           = {10.1103/rvm2-hwx9},
    journal       = {Phys. Rev. D},
    volume        = {112},
    number        = {6},
    pages         = {064049},
    year          = {2025}
}

@article{Lemaitre:1933gd,
    author  = {Lemaitre, G.},
    title   = {{The expanding universe}},
    doi     = {10.1023/A:1018855621348},
    journal = {Annales Soc. Sci. Bruxelles A},
    volume  = {53},
    pages   = {51--85},
    year    = {1933}
}

@article{Tolman:1934za,
    author  = {Tolman, Richard C.},
    title   = {{Effect of inhomogeneity on cosmological models}},
    doi     = {10.1073/pnas.20.3.169},
    journal = {Proc. Nat. Acad. Sci.},
    volume  = {20},
    pages   = {169--176},
    year    = {1934}
}

@article{Bondi:1947fta,
    author  = {Bondi, H.},
    title   = {{Spherically symmetrical models in general relativity}},
    doi     = {10.1093/mnras/107.5-6.410},
    journal = {Mon. Not. Roy. Astron. Soc.},
    volume  = {107},
    pages   = {410--425},
    year    = {1947}
}

@article{Friedman:1922kd,
    author  = {Friedman, A.},
    title   = {{On the Curvature of space}},
    doi     = {10.1007/BF01332580},
    journal = {Z. Phys.},
    volume  = {10},
    pages   = {377--386},
    year    = {1922}
}

@article{Robertson:1935zz,
    author  = {Robertson, H. P.},
    title   = {{Kinematics and World-Structure}},
    doi     = {10.1086/143681},
    journal = {Astrophys. J.},
    volume  = {82},
    pages   = {284--301},
    year    = {1935}
}

@article{Walker:1937qxv,
    author  = {Walker, A. G.},
    title   = {{On Milne's Theory of World-Structure}},
    doi     = {10.1112/plms/s2-42.1.90},
    journal = {Proc. Lond. Math. Soc. s},
    volume  = {2-42},
    number  = {1},
    pages   = {90--127},
    year    = {1937}
}

@article{Ashtekar:2011ni,
    author        = {Ashtekar, Abhay and Singh, Parampreet},
    title         = {{Loop Quantum Cosmology: A Status Report}},
    eprint        = {1108.0893},
    archiveprefix = {arXiv},
    primaryclass  = {gr-qc},
    doi           = {10.1088/0264-9381/28/21/213001},
    journal       = {Class. Quant. Grav.},
    volume        = {28},
    pages         = {213001},
    year          = {2011}
}

@article{Bojowald:2001xe,
    author        = {Bojowald, Martin},
    title         = {{Absence of singularity in loop quantum cosmology}},
    eprint        = {gr-qc/0102069},
    archiveprefix = {arXiv},
    reportnumber  = {CGPG-01-2-1},
    doi           = {10.1103/PhysRevLett.86.5227},
    journal       = {Phys. Rev. Lett.},
    volume        = {86},
    pages         = {5227--5230},
    year          = {2001}
}

@article{Ashtekar:2006rx,
    author        = {Ashtekar, Abhay and Pawlowski, Tomasz and Singh, Parampreet},
    title         = {{Quantum nature of the big bang}},
    eprint        = {gr-qc/0602086},
    archiveprefix = {arXiv},
    reportnumber  = {IGPG-06-2-1},
    doi           = {10.1103/PhysRevLett.96.141301},
    journal       = {Phys. Rev. Lett.},
    volume        = {96},
    pages         = {141301},
    year          = {2006}
}

@article{Ashtekar:2006wn,
    author        = {Ashtekar, Abhay and Pawlowski, Tomasz and Singh, Parampreet},
    title         = {{Quantum Nature of the Big Bang: Improved dynamics}},
    eprint        = {gr-qc/0607039},
    archiveprefix = {arXiv},
    reportnumber  = {IGPG-06-7-2},
    doi           = {10.1103/PhysRevD.74.084003},
    journal       = {Phys. Rev. D},
    volume        = {74},
    pages         = {084003},
    year          = {2006}
}

@article{Modesto:2004xx,
    author        = {Modesto, Leonardo},
    title         = {{Disappearance of black hole singularity in quantum gravity}},
    eprint        = {gr-qc/0407097},
    archiveprefix = {arXiv},
    doi           = {10.1103/PhysRevD.70.124009},
    journal       = {Phys. Rev. D},
    volume        = {70},
    pages         = {124009},
    year          = {2004}
}

@article{Ashtekar:2018lag,
    author        = {Ashtekar, Abhay and Olmedo, Javier and Singh, Parampreet},
    title         = {{Quantum Transfiguration of Kruskal Black Holes}},
    eprint        = {1806.00648},
    archiveprefix = {arXiv},
    primaryclass  = {gr-qc},
    doi           = {10.1103/PhysRevLett.121.241301},
    journal       = {Phys. Rev. Lett.},
    volume        = {121},
    number        = {24},
    pages         = {241301},
    year          = {2018}
}

@article{Gambini:2013ooa,
    author        = {Gambini, Rodolfo and Pullin, Jorge},
    title         = {{Loop quantization of the Schwarzschild black hole}},
    eprint        = {1302.5265},
    archiveprefix = {arXiv},
    primaryclass  = {gr-qc},
    reportnumber  = {LSU-REL-022113, LSU-REL-022113},
    doi           = {10.1103/PhysRevLett.110.211301},
    journal       = {Phys. Rev. Lett.},
    volume        = {110},
    number        = {21},
    pages         = {211301},
    year          = {2013}
}

@article{Bodendorfer:2019cyv,
    author        = {Bodendorfer, Norbert and Mele, Fabio M. and M{\"u}nch, Johannes},
    title         = {{Effective Quantum Extended Spacetime of Polymer Schwarzschild Black Hole}},
    eprint        = {1902.04542},
    archiveprefix = {arXiv},
    primaryclass  = {gr-qc},
    doi           = {10.1088/1361-6382/ab3f16},
    journal       = {Class. Quant. Grav.},
    volume        = {36},
    number        = {19},
    pages         = {195015},
    year          = {2019}
}

@article{Kelly:2020uwj,
    author        = {Kelly, Jarod George and Santacruz, Robert and Wilson-Ewing, Edward},
    title         = {{Effective loop quantum gravity framework for vacuum spherically symmetric spacetimes}},
    eprint        = {2006.09302},
    archiveprefix = {arXiv},
    primaryclass  = {gr-qc},
    doi           = {10.1103/PhysRevD.102.106024},
    journal       = {Phys. Rev. D},
    volume        = {102},
    number        = {10},
    pages         = {106024},
    year          = {2020}
}

@article{Bronnikov:2000vy,
    author        = {Bronnikov, Kirill A.},
    title         = {{Regular magnetic black holes and monopoles from nonlinear electrodynamics}},
    eprint        = {gr-qc/0006014},
    archiveprefix = {arXiv},
    doi           = {10.1103/PhysRevD.63.044005},
    journal       = {Phys. Rev. D},
    volume        = {63},
    pages         = {044005},
    year          = {2001}
}

@article{Ayon-Beato:1998hmi,
    author        = {Ayon-Beato, Eloy and Garcia, Alberto},
    title         = {{Regular black hole in general relativity coupled to nonlinear electrodynamics}},
    eprint        = {gr-qc/9911046},
    archiveprefix = {arXiv},
    doi           = {10.1103/PhysRevLett.80.5056},
    journal       = {Phys. Rev. Lett.},
    volume        = {80},
    pages         = {5056--5059},
    year          = {1998}
}

@article{Dymnikova:1992ux,
    author  = {Dymnikova, I.},
    title   = {{Vacuum nonsingular black hole}},
    doi     = {10.1007/BF00760226},
    journal = {Gen. Rel. Grav.},
    volume  = {24},
    pages   = {235--242},
    year    = {1992}
}
